\documentclass[11pt]{article}
\pdfoutput=1  

\usepackage[T1]{fontenc}
\usepackage[utf8]{inputenc}
\usepackage{mathptmx}
\usepackage[scaled=0.92]{helvet}
\usepackage[margin=1in]{geometry}
\usepackage{microtype}
\usepackage{amsmath,amssymb}
\usepackage{graphicx}
\usepackage{xcolor}
\usepackage{booktabs}
\usepackage{array}
\usepackage{tabularx}
\usepackage{longtable}
\usepackage{makecell}
\usepackage{multirow}
\usepackage{ragged2e}
\usepackage{enumitem}
\usepackage{caption}
\usepackage[most]{tcolorbox}
\usepackage{tikz}
\usepackage{float}
\usetikzlibrary{arrows.meta,positioning,fit,backgrounds,calc,shapes.geometric}
\usepackage[hidelinks,breaklinks]{hyperref}

\setlist{topsep=4pt,itemsep=2pt,parsep=0pt}

\newcolumntype{L}[1]{>{\RaggedRight\arraybackslash}p{#1}}
\newcolumntype{C}[1]{>{\Centering\arraybackslash}p{#1}}

\newcommand{\refitem}[1]{\par\noindent\hangindent=1.6em\hangafter=1 #1\par\addvspace{3pt}}

\definecolor{cFamily}{HTML}{C7E9C0}    
\definecolor{cGreed}{HTML}{FFF2A8}     
\definecolor{cSocProof}{HTML}{D9F0A3}  
\definecolor{cCred}{HTML}{CDEFE6}      
\definecolor{cPersona}{HTML}{F9C6D7}   
\definecolor{cChannel}{HTML}{C6DBEF}   

\definecolor{slate}{HTML}{2B3A48}
\definecolor{cyan}{HTML}{1F9CB5}
\definecolor{amber}{HTML}{E0A030}
\definecolor{paleslate}{HTML}{E7ECF1}
\definecolor{palecyan}{HTML}{DCF1F5}
\definecolor{paleamber}{HTML}{FBEFD6}

\title{\textbf{The Human Vulnerabilities \& Exploits (HVE) Framework}\\[4pt]
\large A Standardized Framework for Cataloguing, Scoring,\\ and Mitigating Human Vulnerabilities}
\author{
  Avichai Ben \\ \small Charm Security 
  \and
  Tom Rahav \\ \small Charm Security 
  \and
  Daniel Illaev \\ \small Charm Security 
  \and
  Aviv Nahon \\ \small Charm Security 
  \and
  Avi Grushka \\ \small Charm Security
}
\date{June 2026}

\begin{document}
\maketitle

\begin{abstract}
\noindent
The cybersecurity community has invested over two decades in building standardized
frameworks, the Common Vulnerabilities and Exposures (CVE) system, the Common
Vulnerability Scoring System (CVSS), and the Common Weakness Enumeration (CWE) to
identify, classify, and remediate threats to digital infrastructure. However, an emerging
body of research reveals that a vast majority of successful cyberattacks exploit not
software flaws, but human behavioral and psychological vulnerabilities. Social
engineering, fraud, and scam attacks, which manipulate human cognition, emotion, and
trust, do not have an equivalent standardized framework. Meanwhile, behavioral science and psychology research has established robust theoretical foundations, such as dual--process theory, prospect theory, social influence frameworks, and visceral state
models, which explain precisely why and how these attacks succeed. This paper
introduces the Human Vulnerabilities \& Exploits (HVE) Framework, a structured
approach for identifying, classifying, and mitigating the behavioral and
psychological vulnerabilities exploited in scams, social engineering, and other
human-centric fraud and attacks, analogous in concept to how CVE helps classify
software vulnerabilities: it provides a shared, machine-readable taxonomy with
structured identifiers, multi-dimensional severity scoring via the Human
Vulnerability Severity Score (HVSS), and actionable remediation guidance through
Human Vulnerability Patches (HVPs). This introduction synthesizes the relevant
literature across cybersecurity standardization, behavioral science, and fraud
defense to establish the theoretical and practical foundations for the HVE
framework, whose architecture and technical specifications are detailed in
subsequent sections.
\end{abstract}

\newpage

\tableofcontents
\bigskip

\section{Introduction}

The cybersecurity community has long relied on standardized frameworks such as the
Common Vulnerabilities and Exposures (CVE) system to manage digital threats.
However, while social engineering is partially addressed in the CVE framework, a
broader analysis of the exploitation of human behavioral and psychological
vulnerabilities one of the main and leading modern attack vectors remains
limited. Research consistently identifies the ``human element'' as a factor in over
60\% of data breaches (Verizon DBIR, 2025; Hancock \& Tessian, 2020), and estimates
suggest that between 84\% and 98\% (Hancock \& Tessian, 2020; IBM Cyber Security
Intelligence Index, 2014) of successful cyber attacks exploit psychological
vulnerabilities rather than purely technical ones. Despite this scale, there is no
standardized equivalent for cataloguing the specific psychological and behavioral
manipulation patterns used in these attacks, leaving a critical blind spot in the
cybersecurity and fraud prevention ecosystem.

This paper introduces the Human Vulnerabilities \& Exploits (HVE) Framework, a novel
framework for human behavioral and psychological vulnerabilities, analogous in
concept to how CVE helps classify and address software vulnerabilities. This
introduction synthesizes literature across cybersecurity standardization including
the CVE ecosystem, which began in 1999, and complementary frameworks like CVSS,
which has evolved through four major versions and behavioral psychology, to
establish the scientific basis for HVE's taxonomy, scoring system, and remediation
model.

\subsection{The CVE Ecosystem: Origins, Architecture, and Lessons}

\subsubsection{Historical Context and Core Architecture}

The Common Vulnerabilities and Exposures system was launched publicly in September
1999 by the MITRE Corporation. It established a universal reference method, a unique
identifier (e.g., CVE-2024-12345) and a standardized description to address the
fragmentation of vulnerability information across proprietary databases.

The modern CVE ecosystem includes the CVE List, the federated network of CVE
Numbering Authorities (CNAs), and the National Vulnerability Database (NVD), which
adds CVSS severity scores and CWE classifications. The Common Vulnerability Scoring
System (CVSS), maintained by FIRST, provides standardized severity ratings on a
0--10 scale, and the Common Weakness Enumeration (CWE) classifies the underlying
patterns of software and hardware weakness. As of April 2026, the CVE ecosystem had
grown to enormous scale: the CVE List contained over 325000 records, with
submissions surging from 28{,}818 in 2023 to 48{,}000 in 2025- a 67\%
increase. The system is referenced by the majority of commercial
vulnerability scanners, is integrated into regulatory compliance frameworks
including PCI-DSS and FedRAMP, and serves as the lingua franca for vulnerability
disclosure between researchers, vendors, and defenders globally.

\subsubsection{Limitations of the CVSS Scoring Model}

CVSS has been criticized for granularity deficiencies, its static design (scores
don't adapt to evolving threats), contextual blindness (assessing inherent severity
but not risk to a specific organization), and poor mapping to emerging technologies
like IoT and AI systems. These shortcomings are critical context for the HVE
Framework's own scoring approach (HVSS).

\subsection{The Human Vulnerability Gap}

\subsubsection{Social Engineering: The Unstandardized Threat}

Social engineering remains limited in its formal cataloguing, despite the Verizon
DBIR identifying the human element in most breaches (Verizon DBIR, 2025). The
industry's primary response the ``human firewall'' concept delivered via security
awareness training is mostly reactive, lacks standardized vocabulary and scoring,
and doesn't address the rapidly growing category of scams and human-centric fraud
attacks targeting consumers, users, and end-customers outside corporate
environments.

\subsubsection{Academic Frameworks for Human Vulnerability}

Academic research has sought to systematize this threat, identifying how attackers
exploit cognitive biases and proposing taxonomies based on dimensions like
environment, approach, and delivery medium. Researchers have suggested treating
human cognition as an ``attack surface'' amenable to rigorous analysis,
demonstrating that psychological exploitation can be systematically catalogued but
none has produced an operationally deployable framework comparable to CVE.

\subsection{Behavioral Psychology of Fraud and Scam Victimization}

\subsubsection{Cognitive Foundations: Dual-Process Theory and Prospect Theory}

Dual-process theory (System 1: quick, intuitive; System 2: slow, analytical)
explains that System 1's reliance on heuristic shortcuts makes it vulnerable to
manipulation. Scammers systematically exploit this by activating System 1 through
emotional triggers (fear, urgency, and others) while preventing System 2 engagement.
Prospect theory complements this: individuals often experience potential losses more
intensely than equivalent gains, which scammers exploit by framing communications to
activate loss-aversion circuits, making victims more risk-seeking.

\subsubsection{Emotional and Visceral Mechanisms}

Robert Cialdini's seven principles of influence (Cialdini, 2001) Reciprocity,
Commitment and Consistency, Social Proof, Authority, Liking, Scarcity, and
Unity are the most widely cited framework in persuasion-based fraud. Complementary
frameworks such as those identifying principles like Distraction, Social
Compliance, Herd, and Time reveal a consistent set of exploitable psychological
mechanisms that can be systematically catalogued and defended against.

George Loewenstein's hot--cold empathy gap demonstrates that individuals in a calm
(``cold'') state underestimate the influence of visceral drives and emotions in an
aroused (``hot'') state. Fraud and scams that induce extreme emotional states are
most likely to override rational self-protective behavior. Victims often enter a
``spell'' state characterized by panic and impaired reasoning, which requires
interventions that introduce cognitive friction to allow System 2 to re-engage.

\subsubsection{Demographic Vulnerabilities and Individual Differences}

Research identifies age-related and personality-based risk factors. Mild cognitive
impairment is linked to increased scam vulnerability, and the neurobiological
mechanisms underlying age-related vulnerability include reduced cortical volume and
declines in processing speed and memory. Personality traits like impulsivity, high
trust, and sensation-seeking are also associated with elevated fraud
risk emphasizing that interventions must address psychological vulnerabilities,
not just knowledge.

\subsubsection{Grooming, Relationship Exploitation, and AI-Amplified Fraud}

Long-con operations, like romance scams, systematically exploit interpersonal
communication frameworks to construct deep emotional bonds before financial
requests. Generative AI has fundamentally altered the threat landscape: AI-generated
phishing emails have achieved an up to 4.5$\times$ effectiveness multiplier (Heiding
et al., 2024) over traditional phishing, and deepfake voice cloning lowers the
barrier to deploying psychologically powerful attacks.

\subsection{Behavioral Interventions and Their Limitations}

One of the most effective behavioral interventions against active scam attempts is
``cognitive friction'' introducing delays, verification steps, or reframing
prompts to allow System 2 to re-engage. Emerging research on real-time scam
intervention emphasizes matching the victim's emotional state before attempting to
redirect behavior, a technique known as ``pacing and leading.'' Rather than directly
contradicting the victim's beliefs (which triggers confirmation bias and rejection),
effective interventions first validate the victim's emotional experience, then
gradually introduce verification steps framed as helpful rather than obstructive.
This avoids the backfire effect documented in persuasion research, where direct
contradiction of strongly held beliefs paradoxically strengthens those beliefs.

Multiple studies show that knowledge-based fraud prevention educational campaigns,
awareness training, and informational materials is necessary but insufficient for
reducing victimization. The hot--cold empathy gap explains why: knowledge acquired
in a ``cold'' state fails to transfer to behavior in the ``hot'' emotional state
induced by an active scam. Effective prevention must therefore combine education
(cold-state preparation) with real-time behavioral interventions (hot-state
support) a design principle that directly informs the HVE Framework's Human
Vulnerability Patch (HVP) system.

\subsection{Toward the HVE Framework: Bridging the Gap}

The cybersecurity community has mature frameworks for technical vulnerabilities and a
rich body of behavioral science on human exploitation, yet no operationally
deployable framework bridges these domains. The Human Vulnerabilities \& Exploits
(HVE) Framework is explicitly designed to fill this gap. Where CVE asks ``What
software flaw exists?'', HVE asks ``What human vulnerability did the attacker
exploit, and how can we patch it?''

The HVE framework introduces several novel contributions:
\begin{itemize}
  \item \textbf{3-Layer DNA Model:} Decomposes attacks into Manipulation Family (the
  scenario), Social Engineering Tactics (SET) (the psychological lever), and
  Persona (the impersonated identity).
  \item \textbf{Structured Record Schema:} Captures attack vectors, predisposing
  factors, and emotional triggers, bridging cybersecurity and behavioral science.
  \item \textbf{Human Vulnerability Severity Score (HVSS):} Addresses CVSS
  limitations through Base Metrics (attack properties) and Context Metrics (victim
  state and susceptibility).
  \item \textbf{Spell Intensity Classification:} Operationalizes
  visceral states (Low/Medium/High) to determine the appropriate intervention mode.
  \item \textbf{Human Vulnerability Patches (HVPs):} Translates cognitive friction
  research into machine-readable, agent-deployable intervention bundles.
\end{itemize}

The remainder of the paper proceeds as follows: Section~\ref{sec:problem} motivates
the gap; Section~\ref{sec:related} positions HVE against related work;
Section~\ref{sec:hve} details the HVE Framework's architecture, including the
21-family taxonomy, the 3-Layer DNA, the HVE \& HWE registries and the HVSS \& HVP
concepts, culminating in a worked example; Section~\ref{sec:agent} sketches the
real-time agent deployment pipeline; and Section~\ref{sec:usecases} outlines key HVE
Framework use cases.

\section{Problem Statement and Motivation}\label{sec:problem}

The problem motivating the HVE framework is a critical operational liability: while
the cybersecurity community has built mature infrastructure (like CVE) for technical
flaws, it has a limited framework for the psychological vulnerabilities that
attackers exploit at scale. This gap is defined by a lack of standardization
comparable to software's CVE/CVSS/CWE ecosystem: defenders lack a shared identifier
space, a multi-dimensional severity score, and a structured remediation pathway for
distinct human manipulation patterns.

Existing partial solutions are insufficient:
\begin{itemize}
  \item Security awareness training is focused on employees and does not address the
  broader end-customer and end-user base; it also provides limited value because
  knowledge acquired in a calm state does not transfer to the ``hot'' emotional
  state induced by an active scam.
  \item MITRE F3 (Fight Fraud Framework), launched in April 2026, is important
  progress in the fraud prevention space, but is mostly attacker-centric, lacking
  vocabulary for victim psychology, severity scoring, and real-time intervention.
  \item Academic taxonomies are descriptive but not operationally deployable or
  machine-readable.
\end{itemize}

The rapid maturation of AI-driven social engineering is closing the window for an
unstandardized response: AI-generated phishing has shown a 4.5$\times$ effectiveness
multiplier (Heiding et al., 2024), and deepfakes are increasing the volume and
sophistication of personalized, emotional long-cons that previously required skilled
human operators.

A solution must provide a globally unique identifier, decompose attacks into
reusable elements, include a multi-dimensional severity score, be grounded in
behavioral literature, and bind vulnerability descriptions to agent-deployable
intervention bundles (HVPs). HVE is designed to satisfy these requirements, shifting
the focus from ``what the attacker did'' to ``what human vulnerability the attacker
exploited and how it can be patched.''

\section{Related Work and Positioning vs.\ Existing Frameworks}\label{sec:related}

The HVE framework draws on three streams of prior work: consumer-facing scam
taxonomies; cybersecurity standardization frameworks (CVE, CVSS, CWE) and adversary
models (ATT\&CK, F3); and the behavioral science literature on social influence,
dual-process cognition, and grooming. Each contributes essential building blocks yet
leaves a structural gap HVE is explicitly designed to close. This section situates
HVE within that landscape and answers the reviewer's question of ``why HVE and not
X?'' for every plausible X.

\subsection{Existing Scam and Fraud Taxonomies}

Existing scam and fraud taxonomies such as the Federal Reserve ScamClassifier
(2024) and FraudClassifier (2020), Better Business Bureau (BBB, 2024) Scam Tracker,
and the more rigorous ASU/FINRA taxonomy (Shang et al., 2025; Beals et al.,
2015) classify the cover story (e.g., ``utility shutoff,'' ``IRS impersonation'')
but fail to model the psychological mechanism that makes the attack succeed in real
time.

The BBB and Federal Reserve taxonomies lack severity scoring or a model of victim
manipulation. While the hierarchical ASU/FINRA taxonomy is more systematic, all miss
three critical signal classes necessary for real-time defense:
\begin{itemize}
    \item \textbf{Grooming Depth: }Distinguishes between flash attacks and long-con exploitation.
    \item \textbf{Spell Intensity: }Measures the victim’s real-time visceral state to determine intervention.
    \item \textbf{Composability: }Enables decomposition of complex attacks into Family(Scenario) × SET × Persona triples.
\end{itemize}

HVE composes with these existing taxonomies by mapping its Family field to their
labels while adding the psychological exploit layer (SET and Persona) to create an
actionable vulnerability record.

\subsection{Cybersecurity Standards: The CVE/CWE Analogy}

HVE is explicitly modeled on the architecture of the CVE ecosystem, borrowing its
foundational design choices: a unique, immutable identifier (HVE-ID), a structured
JSON schema, and a federated allocation registry. The Human Vulnerability Scoring
System (HVSS) is a CVSS-style 0--10 composite that combines a Base component
(intrinsic exploit potency) with a Context component (deployment-specific
modifiers). The taxonomic layer is modeled on CWE, classifying underlying weakness
patterns into parallel family groups (e.g., Romance, Authority). The key conceptual
difference is the object of description: CVE describes technical artifacts, while HVE
describes human psychological states and provides intervention bundles (HVPs)
instead of code changes.

\subsection{MITRE ATT\&CK and the Fight Fraud Framework (F3)}

HVE maintains deliberate structural parallels with MITRE ATT\&CK and the Fight Fraud
Framework (F3) to ensure interoperability for instance, an ATT\&CK Mitigation is
analogous to an HVE Human Vulnerability Patch (HVP). However, ATT\&CK and F3 model
the attacker's behavior and infrastructure (TTPs), while HVE models the victim's
cognition and psychological state and its exploitation the reason the attack worked
at that moment. For example, F3 records that an attacker performed ``vishing,'' while
HVE decomposes the action into its constituent psychological mechanisms (e.g.,
exploiting Authority bias or triggering Fear), providing the missing vocabulary for
victim-side modeling, scoring, and intervention. HVE thus describes the cognitive
exploit chain that TTPs run on.

\begin{figure}[ht]
\centering
\includegraphics[width=\linewidth]{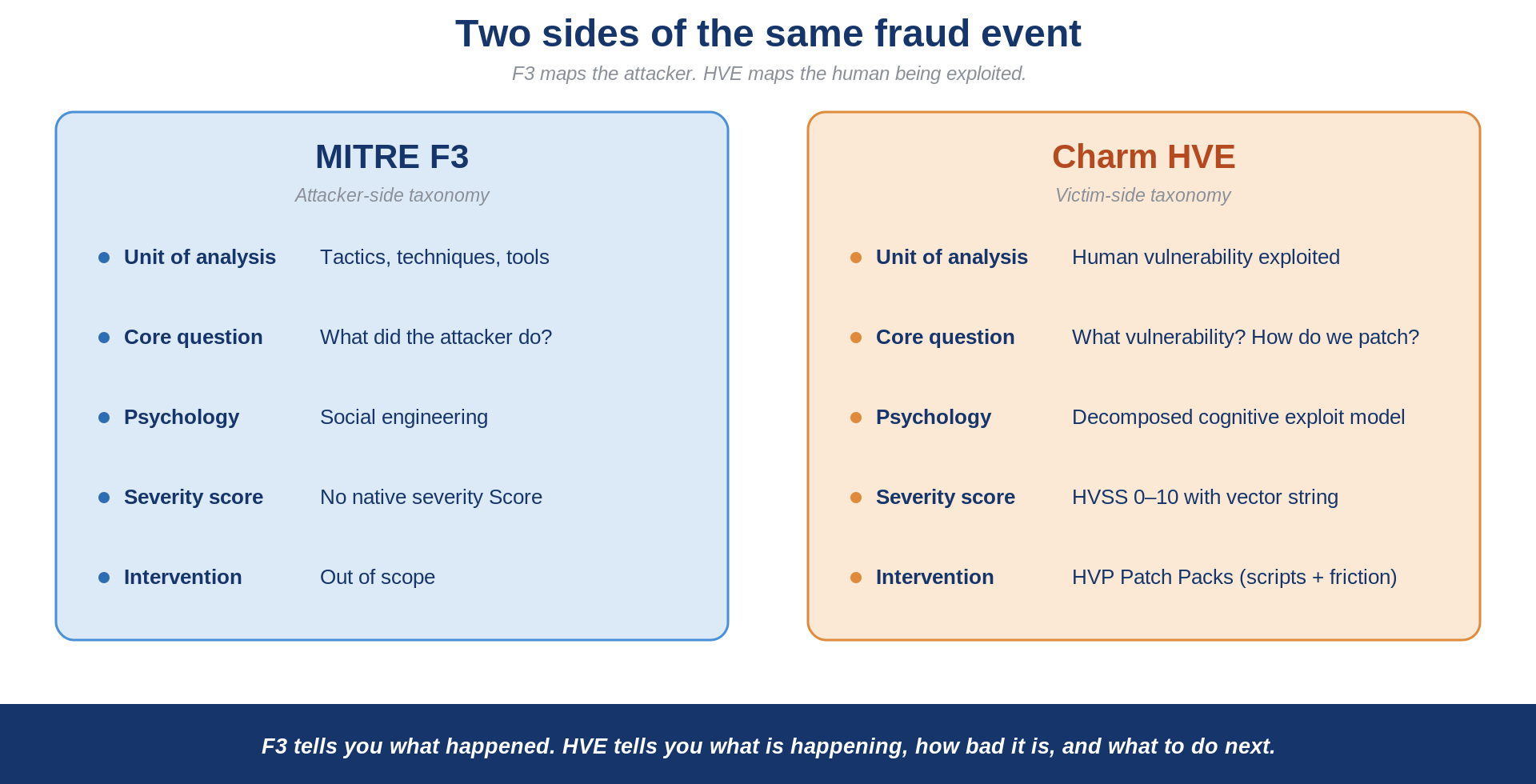}
\caption{Two sides of the same fraud event. F3 describes the attacker's behavior; HVE describes the victim-side vulnerability exploited, with severity scoring (HVSS) and intervention bundles (HVPs). (Vector reconstruction of the source figure.)}
\label{fig:anatomy}
\end{figure}

\subsection{Behavioral Psychology Literature}

HVE's psychological layer is scientifically grounded in behavioral science
literature on social influence, dual-process cognition, visceral emotion, and
grooming. Every machine-readable field operationalizes an established construct:
\begin{itemize}
  \item \textbf{Technique taxonomy} based on Cialdini's seven principles of
  influence (Reciprocity, Authority, Scarcity, etc.), extended by frameworks from
  Stajano and Gragg.
  \item \textbf{Grooming Depth} (Flash, Multistep, Long-con) uses crime-script
  analysis to classify temporal horizons, which determines the intervention class.
  \item \textbf{Spell Intensity} (Low/Medium/High) operationalizes Loewenstein's
  hot--cold empathy gap, ensuring the Human Vulnerability Patch (HVP) is calibrated
  to the victim's visceral state (logic at Low, friction at Medium, pacing and
  leading at High).
\end{itemize}

To the best of our knowledge, HVE is the first framework to operationalize Cialdini,
Loewenstein, and the grooming literature into a single machine-readable form for
real-time AI agent consumption.

\subsection{Positioning}

\begin{table}[ht]
\centering
\caption{Positioning of HVE relative to existing cybersecurity and fraud frameworks.}
\label{tab:positioning}
\small
\begin{tabularx}{\linewidth}{@{}L{0.20\linewidth} L{0.40\linewidth} L{0.36\linewidth}@{}}
\toprule
\textbf{Dimension} & \textbf{CVE / CVSS / CWE \& ATT\&CK / F3} & \textbf{HVE Framework}\\
\midrule
Severity scoring & CVSS (CVE), none (F3) & HVSS 0--10 with Base + Context\\
Psychological layer & Absent & Social Engineering Tactics, Grooming Depth, Spell Intensity\\
Composability & Technique $\times$ Mitigation pairing (CVE) & Taxonomy $\times$ Tactics $\times$ Persona DNA + HVP\\
Real-time agent integration & Retrospective (F3), SIEM-oriented (ATT\&CK) & Vector strings + 4-step agent workflow + HVP Patch Packs\\
\bottomrule
\end{tabularx}
\end{table}

The position HVE occupies is distinctive in a precise sense. It is not a new
framework for technical artifacts CVE, CWE, and CVSS occupy that ground, and HVE
inherits their conventions. It is not a new adversary behavior model ATT\&CK and F3
already provide that, and HVE mirrors their Technique/Mitigation parallel at the
cognitive layer. It is, instead, the missing victim-side companion to all of them:
the standardized description of the cognitive exploit chain underneath every social
engineering attack.

Types of frauds and scams are an effect; manipulation patterns are the cause. HVE
organizes the cognitive exploits themselves the layer at which severity is
determined and interventions succeed.

\section{HVE: A Human-Centric Fraud Universal Language}\label{sec:hve}

The HVE framework aims to create the missing universal language for human-targeted
fraud by providing a shared, machine-readable taxonomy with unique identifiers and
severity scoring, analogous to the CVE system for software.

\subsection{HVE Family Taxonomy}

The Family field uses a 21-value controlled vocabulary to capture the shared trust
model an attacker exploits. This taxonomy, derived from empirical case reports and
academic literature, enables standardized classification, search, and intervention
selection across the HVE lifecycle.

\paragraph{Design criteria.} A candidate family was admitted only if it satisfied all
three of the following.
\begin{itemize}
  \item \textbf{Operational distinguishability.} The family must correspond to a
  distinctive cluster of responses. Two families that would trigger the same HVP, the
  same verification flow, and the same demographic-aware deployment are one family
  with sub-variants, not two families.
  \item \textbf{Representational coverage.} The union of families should cover the
  empirical case distribution with a low residual rate. Residuals, when they arise,
  are handled at the record level rather than by introducing further families.
  \item \textbf{Psychological coherence.} Within a family, the dominant Social
  Engineering Tactics and Persona archetypes must form a coherent cluster. AUTH is
  coherent because the dominant tactics are Authority + Fear + Urgency and the
  persona is institutional; ROM is coherent because the dominant tactics are Liking +
  Commitment and the persona is an intimate partner.
\end{itemize}
The Family field uses a controlled vocabulary of 21 codes; the full list, including
detailed descriptions, is presented in Appendix~\ref{app:taxonomy}.

\paragraph{Principled overlaps.} Two overlaps in the taxonomy deserve specific
comment.

The romance-investment hybrid, widely called ``pig butchering,'' is treated as two
linked records, one ROM and one INVT, rather than as a separate family. We make this
choice because the two layers are temporally separable (relationship grooming
precedes investment-platform engagement), target different psychological exploits
(Liking and Commitment in the romance phase; Scarcity and Authority in the investment
phase), and are, on the evidence of Wang and Kelsay (2025) and Asyali et al.\ (2026),
increasingly observed alongside non-hybrid forms. Representing them as linked records
preserves both the joint pattern and the ability to analyze each half independently.

BIZ and EXEC are the only enterprise-targeted families; they are retained in the
otherwise consumer-oriented taxonomy because the same psychological mechanisms are
operative, but their associated HVPs differ substantially. They target procurement
controls and out-of-band verification of payment changes rather than the
single-victim pacing-and-leading scripts that dominate the consumer families.

\paragraph{Operational use.} The Family field is consumed at three distinct points in
the HVE lifecycle. In candidate matching, family membership is inferred jointly with
other features from the conversation digest during vector-similarity search; the
inferred family then constrains which specific records are scored as matches,
narrowing the search space by an order of magnitude in the typical case. In
intervention selection, HVPs are indexed by family as well as by specific HVE-ID,
allowing fallback to a family-level patch when a specific record match is
low-confidence the analogue of falling back from a specific CVE patch to a
CWE-level mitigation. In enterprise reporting, partners consume HVE telemetry at the
family level, with the 21-family $\times$ channel $\times$ demographic-band cube
serving as the canonical slice for heatmaps and patch-effectiveness measurement
(Section~\ref{sec:usecases}).

Extensibility is built in along two paths. Sub-family refinement introduces finer
codes within a family (AUTH-Police vs.\ AUTH-Court vs.\ AUTH-Regulator) at the record
level without changing the taxonomy, and this is the preferred path for most observed
evolution. New-family introduction is reserved for emerging clusters that cannot be
represented as sub-families of existing ones; historical records are not renumbered,
and the year component of the HVE-ID captures the taxonomic epoch.

\subsection{The 3-Layer DNA}

At the center of every HVE record sits a three-field decomposition:
\textbf{Manipulation Family}, \textbf{Social Engineering Tactics}, and \textbf{Persona
Archetype}. The three layers are epistemically distinct. Manipulation Family is the
story, Persona Archetype is the storyteller, and Social Engineering Tactics are the
levers that make the story work. Together they constitute the minimal description from
which the rest of the record can be reasoned about, and the decomposition is a direct
application of the third design principle below: a scam is a composition of reusable
elements, and the DNA names those elements.

\textbf{Manipulation Family} is the shared trust model the attacker exploits the
institutional, relational, or commercial relationship that gives the cover story its
credibility. A grandson in a holding cell, an account compromised and needing a
transfer to a ``safe'' account, back taxes demanding immediate payment, and a broker
with guaranteed returns each anchor to a single Family: FAM, BANK, GOV, and INVT
respectively. Family is what a victim would name if asked who the message appeared to
come from. Pretexts are reusable across families: the Bail/Arrest pretext appears
within AUTH, FAM, and sometimes GOV records, carried as a sub-family at the record
level rather than as a standalone field.

\textbf{Social Engineering Tactics} are the psychological levers that suppress
analytical reasoning and drive compliance. In dual-process terms, they are the
System-1 activations that block System-2 engagement. The full Tactics vocabulary is
curated by Charm Security from leading academic frameworks (e.g., Cialdini 2001 on
social influence; Gragg 2003 on social engineering triggers); overlapping constructs
across sources have been reconciled into a single list, and sub-tactics added where
the empirical literature supports finer distinction (e.g., Visceral Influence
decomposes into Greed, Excitement, Fear, and Sexual desire).

\textbf{Persona Archetype} is the trust-relationship the attacker is impersonating,
drawn from a controlled vocabulary of eight values: Authority Figure (police officer,
bank fraud analyst, tax official, judge, lawyer, doctor), Familiar Person (grandson,
family member, boss, colleague), The Professional / Expert (financial guru,
investment advisor, recruiter, real-estate agent), Benevolent Helper (tech support
agent, utility rep, account recovery specialist, healthcare provider), Romantic
Persuader, Help-Seeker, Gatekeeper, and Intimidator. Persona Archetype captures the
trust model the attacker is exploiting and determines both the verification path
encoded in the HVP and the Match-Score component of HVSS, since archetypes align
differentially with demographic predisposing factors (Boyle et al., 2014; Koning and
Junger, 2024).

We considered both finer and coarser decompositions and rejected them on design
grounds: finer decompositions fragment the psychological lever across fields that
annotators cannot reliably separate, while coarser ones collapse the verification-path
information that the associated HVP depends on.

\paragraph{Worked examples.} The grandparent-bail scam comprises Manipulation Family
AUTH (Bail/Arrest sub-family), Authority + Urgency + Fear + Complicity tactics, and an
Authority Figure (Police) persona, necessitating callback verification friction. In
contrast, the pig-butchering investment scam combines Manipulation Family INVT with
Liking + Similarity + Greed tactics and a Professional / Romantic Persuader persona,
requiring investment-platform validation and a related-family link to ROM due to the
distinct temporal and psychological phases of the grooming process.

\subsection{Design Principles}

Seven principles govern the framework. They are not independent; the schema is the
product of their joint satisfaction.
\begin{enumerate}
  \item \textbf{Interoperability over novelty.} Where CVE already solves a problem,
  HVE reuses its solution.
  \item \textbf{Minimal-but-complete records.} A field is admitted only if it
  identifies the vulnerability, carries decision-relevant information for a defender,
  or is required to compute HVSS or select an HVP.
  \item \textbf{Decomposition over monolithic description.} A scam is a composition of
  reusable elements, not an atomic event.
  \item \textbf{Psychological grounding.} Every field that describes a human factor
  resolves to a construct in the peer-reviewed behavioral literature.
  \item \textbf{Agent-deployability.} The schema is designed for real-time consumption
  by AI agents, not only for retrospective analysis.
  \item \textbf{Lifecycle and governance fitness.} Governance resilience is built into
  the schema rather than bolted on later.
  \item \textbf{Separation of attack and victim.} The attack pattern (HVE) and the
  victim's vulnerability profile (HWE) are independent registries that combine at
  scoring time (HVSS), not at record-creation time.
\end{enumerate}

\subsection{Record Schema}

The record extends the 3-Layer DNA with the additional fields required to make it
actionable, organized into four blocks. Each block is discussed below together with
the principle that governs its design.

\paragraph{Core identity.} Each record carries a globally unique HVE-ID of the form
\texttt{HVE-YYYY-<FAMILY>-NNNN}. The format is a direct inheritance from CVE
(Principle 1). The year component gives temporal anchoring for trend analysis; the
family component embeds CWE-like taxonomic information into the identifier itself, so
that \texttt{HVE-2026-AUTH-0042} is recognizably an Authority record without a
database lookup. The four-digit sequence matches CVE width for visual symmetry in
co-displayed records. Family is kept as a separate field even though it is redundant
with the identifier, because it is the natural join key for grouped trend analysis and
because it allows cross-references without parsing identifier strings.

\paragraph{Attack vector.} The Exposure Channel field enumerates the medium of
delivery: Phone, SMS, Messaging App, Email, Marketplace, Dating App, Social DM,
In-person. Channel is separated from Manipulation Family and Persona because
channel-specific detection and friction mechanisms are distinct: a Bank Fraud Analyst
persona in the Phone channel triggers a callback-verification patch, while the same
persona in the Email channel triggers a DMARC-and-link-analysis patch. The field is
small, closed, and cheap to populate (Principle 2), and it earns its place by letting
detection engineering be configured per channel.

\paragraph{Psychological exploit details.} This block is the HVE-specific
contribution the fields that have no CVE counterpart and that F3 explicitly lacks.
It is the direct application of Principle 4: every field resolves to a construct in
the peer-reviewed behavioral literature.

SETs enumerate the social-engineering methods in play (Authority Impersonation,
Altercasting, Foot-in-the-Door, Pretexting, Commitment Escalation), drawn from
Cialdini (2001), Stajano and Wilson (2011), and Gragg (2003), and are multi-valued
because real attacks layer techniques: a pig-butchering conversation typically
combines Liking, Commitment Escalation, and Scarcity over weeks (Asyali et al., 2026;
Wang and Kelsay, 2025).

Grooming Depth is categorized as Flash (minutes), Multi-step (days to weeks), or
Long-con (months); we include it both as a descriptive field and as a severity input,
on the hypothesis supported by Wang and Kelsay (2025) and Asyali et al.\
(2026) that longer grooming correlates with stronger emotional investment and
greater resistance to direct contradiction.

\paragraph{Mitigation.} The Mitigation block is what makes a record actionable rather
than merely informational. Where CVE records describe a vulnerability and leave
remediation to downstream patch-management systems, HVE binds each vulnerability
record to a specific intervention bundle at the schema level. This tight
coupling psychological record to deployable patch is, to our knowledge, the first
time it has been attempted in a fraud-defense standard. Three fields realize the
coupling.

Indicators \& Evidence enumerate the observable cues that signal exploitation,
including linguistic markers (bail, police, wire, urgent), payment-rail indicators
(gift cards, crypto, mule accounts), and behavioral signals (long-hold phone sessions,
refusal to disconnect), each carrying a required confidence level so that downstream
detection systems can calibrate their false-positive tolerance.

The Associated Patch Pack is a typed pointer to the HVP record containing the
intervention bundle. The Response Playbook enumerates recommended behaviors across the
Protection, Prevention, and Remediation lifecycle, ensuring the record covers the full
victim timeline rather than terminating at the point of loss.

\paragraph{A populated record.} Table~\ref{tab:record} shows the grandparent-bail
record introduced in Section~\ref{sec:hve} with all four schema blocks populated. The
record is abbreviated for readability; the full taxonomy is given in
Appendix~\ref{app:taxonomy}.

\begin{table}[ht]
\centering
\caption{A populated HVE record (grandparent-bail scenario), abbreviated.}
\label{tab:record}
\small
\begin{tabularx}{\linewidth}{@{}L{0.16\linewidth} L{0.24\linewidth} X@{}}
\toprule
\textbf{Block} & \textbf{Field} & \textbf{Value}\\
\midrule
\multirow{2}{*}{\textbf{Core Identity}} & HVE-ID & \texttt{HVE-2026-AUTH-0042}\\
 & Name & Authority Impersonation: Bail Scenario (Grandparent Scam)\\
\midrule
\multirow{3}{*}{\textbf{3-Layer DNA}} & Family & AUTH\\
 & Persona Archetype & Authority Figure\\
 & Social Engineering Tactics & Authority, Urgency, Fear, Complicity\\
\midrule
\textbf{Grooming} & Grooming Depth & Flash (single interaction)\\
\midrule
\textbf{Attack Vector} & Exposure Channel & Phone (Voice)\\
\midrule
\multirow{3}{*}{\textbf{Mitigation}} & Indicators \& Evidence & Keywords ``bail'', ``police'', ``wire''; demand for secrecy; refusal to disconnect; pressure to act before verification\\
 & Associated Patch Pack & \texttt{HVP-AUTH-0042-Verify}\\
 & Response Playbook & \textit{Protection:} flag inbound authority-persona calls. \textit{Prevention:} apply friction delay, pacing-and-leading, request case-number verification via official channel. \textit{Remediation:} capture caller ID and transfer destination, connect victim with emotional-support resources.\\
\bottomrule
\end{tabularx}
\end{table}

\subsection{Vector String Fingerprint}

Each record carries a compact vector string that projects its key structured fields
onto a single line. For the grandparent-bail record:

\begin{tcolorbox}[colback=paleslate,colframe=slate,boxrule=0.4pt,arc=2pt,
  left=6pt,right=6pt,top=4pt,bottom=4pt,fontupper=\ttfamily\footnotesize]
HVEV1: FAMILY=AUTH; PERS=AuthorityFigure/Police;\\
AV=Phone; SE=Authority+Urgency+VI(Fear)+Complicity; GD=Flash
\end{tcolorbox}

The JSON record remains authoritative; the vector is a lossy projection chosen to
support vector-similarity matching during candidate selection and to travel inside
transaction-level telemetry without additional encoding.

\subsection{HWE Record (Human Weakness Enumeration)}

The victim side of every attack is catalogued in its own registry. ``Weakness'' is
used here in the CWE sense: a structured, non-judgmental classification of
susceptibility conditions that may be exploited in a given context, not a label of
personal deficiency or blame. An HWE record answers the question of which contextual
or susceptibility factors may be exploitable in a given situation, and is independent
of any specific HVE: the same weakness can be exploited by many different attack
patterns, and the same attack pattern can target many different weaknesses. Separating
attack from victim is Principle 7 of the framework (``separation of attack and
victim'') and enables combinatorial matching at HVSS scoring time. Each HWE carries a
globally unique identifier of the form \texttt{HWE-YYYY-NNNN}, with its own lifecycle
and governance distinct from HVE.

Predisposing Factors capture victim-targeting criteria (age band, cognitive-reflection
proxies, social isolation, financial-literacy band, recent-event context), grounded in
Boyle et al.\ (2014), James et al.\ (2015), and Koning et al. (2024); they drive
the Match-Score component of HVSS.

\paragraph{HWE record schema.} The HWE record carries a Core Identity block (HWE-ID,
Name) and Predisposing Factors.

\begin{table}[ht]
\centering
\caption{HWE record schema.}
\label{tab:hwe-schema}
\small
\begin{tabularx}{\linewidth}{@{}L{0.18\linewidth} L{0.20\linewidth} X@{}}
\toprule
\textbf{Block} & \textbf{Field} & \textbf{Value}\\
\midrule
\multirow{2}{*}{\textbf{Core Identity}} & HWE-ID & \texttt{HWE-YYYY-NNNN}\\
 & Name & Human-readable label (e.g., ``Social Isolation with Authority Bias'')\\
\midrule
\textbf{Predisposing Factors} & Factor(s) & Multi-valued, from controlled vocabulary\\
\bottomrule
\end{tabularx}
\end{table}

Predisposing Factors are contextual, situational, or behavioral indicators such as
social isolation or recent life events that may increase susceptibility in a
specific interaction and are known to be leveraged by attackers for targeting or
manipulation. These factors can be used to calibrate protective interventions, not to
label individuals or support adverse decisions.

\paragraph{A populated HWE record.} The grandparent-bail HVE
(\texttt{HVE-2026-AUTH-0042}) pairs with \texttt{HWE-2026-00187}, ``Social Isolation
with Authority Bias.'' Predisposing Factors: Social Isolation. In a worked record,
HVSS would summarize this exploit as high-severity, triggering a High-intensity
intervention tier. A different HWE paired with the same HVE for example a financially
strained working-age adult without elevated Authority Bias, would score lower on
Context and select a different HVP variant.

\subsection{Human Vulnerability Severity Score (HVSS)}

The Human Vulnerability Severity Score (HVSS) is a multi-dimensional severity rating,
on a 0--10 scale, designed to be conceptually similar to the Common Vulnerability
Scoring System (CVSS) but applied to human vulnerabilities. It uniquely incorporates
the full context of an attack by factoring in both the attack's intrinsic properties
(from the HVE record) and the victim's psychological status (from the HWE record). The
HVSS outputs a dynamic, contextual score for a specific identified social engineering
technique as applied to a specific victim profile.

\subsection{Human Vulnerability Patches (HVPs)}

The Human Vulnerability Patches (HVPs) are machine-readable intervention bundles linked
directly to the HVSS. They translate cognitive friction research and pacing-and-leading
strategies into deployable actions. An HVP is selected based on the HVSS score, which
derives its inputs from the HVE (attack) and HWE (victim) records. It is calibrated to
the victim's computed Spell Intensity (Low, Medium, or High) to ensure the intervention
is the most fitting response to the victim's psychological state.

\subsection{Trade-offs and Non-goals}

Three trade-offs are worth stating explicitly. First, controlled vocabularies replace
free text for fields such as Manipulation Family, Social Engineering Tactics, Persona
Archetype, and Exposure Channel. This sacrifices expressive fidelity in exchange for
machine-comparability and stable classifier inputs; vocabulary extension is handled
through governance, not ad hoc escape hatches. Second, records are intended to describe
vulnerability patterns rather than individual victim incidents: an
Authority-Impersonation bail scam targeting elderly victims via phone is one record,
not ten thousand. This matches the CVE convention; individual-incident data is the
concern of downstream telemetry systems that consume HVE records as input. Third, HVE
prescribes a scoring system and intervention bundles but does not prescribe the
enforcement thresholds at which a partner should block a transaction, drop a call, or
suspend an account. Those thresholds depend on risk appetite and regulatory
environment, and attempting to fix them in the framework would both misallocate
authority and depress adoption.

Two non-goals are equally worth stating. HVE does not model attacker reconnaissance,
resource development, or monetization infrastructure; that layer is well-covered by F3
and replicating it would violate the interoperability principle. And HVE does not
attempt to encode novel psychological theories; every human-factor field resolves to an
existing construct in the behavioral literature, so that empirical validation can
proceed against a shared theoretical base.

\subsection{HVE Example Walkthrough}\label{sec:walkthrough}

For the purposes of the walkthrough we demonstrate the framework on a case brought
from the BBB (Better Business Bureau).

\begin{figure}[ht]
\centering
\includegraphics[width=\linewidth]{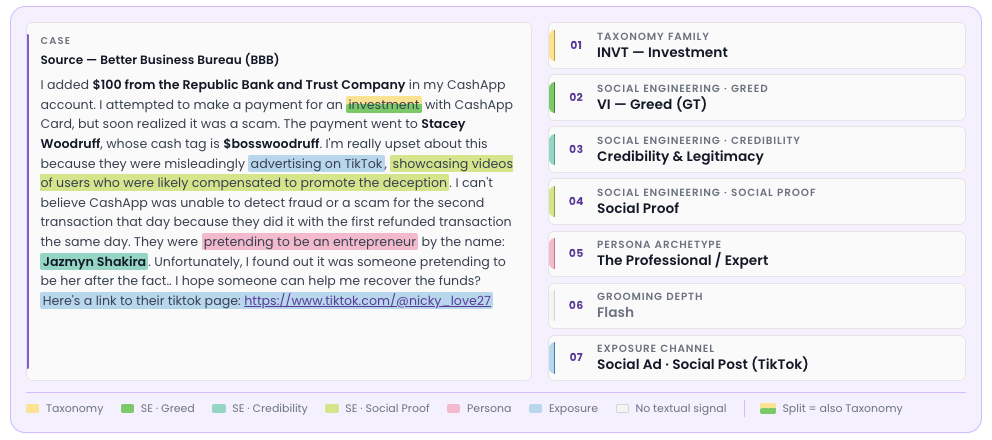}
\caption{HVE feature-axis labelling of a BBB scam case (Section~\ref{sec:walkthrough}).
Each colour maps a textual span to a single HVE feature axis. (Vector reconstruction of
the source figure; case text paraphrased to preserve the highlighted spans.)}
\label{fig:walkthrough}
\end{figure}

We label this case by mapping each highlighted span to a single HVE feature axis, with
the colours carrying the mapping (Figure~\ref{fig:walkthrough}). The word
``investment'' is split-coloured yellow over green because it evidences two axes at
once: the Manipulation Family is INVT (Investment) and the first social-engineering factor
is Greed (VI), the lure of returns. The lime-green span on ``showcasing videos of users
\dots\ to promote the deception'' records Social Proof fabricated testimonials that
borrow credibility from a fake crowd. The mint-green highlight on ``Jazmyn Shakira''
records Credibility \& Legitimacy: a real, locatable named identity is what makes the
impersonation persuasive. The pink highlight on ``pretending to be an entrepreneur''
isolates the Persona Archetype The Professional / Expert to the act of role-claiming
itself, not to the surrounding narrative. The blue spans on ``advertising on TikTok''
and the linked profile fix the Exposure Channel as Social Ad / Social Post (TikTok).
Grooming Depth is Flash, since the interaction collapses into a single transaction with
no sustained rapport-building; that axis carries no textual signal and is left
unhighlighted.

\section{Agent Integration and Real-Time Deployment}\label{sec:agent}

The preceding sections have established the HVE, HWE, HVSS, and HVP registries as
structured artifacts. This section describes how those artifacts become operational:
how an analysis AI agent, processing a live conversation, applies all four registries
in sequence to produce a structured assessment and a targeted intervention.

\subsection{The Four-Registry Analysis Pipeline}

\subsubsection{Attack Classification (HVE)}

The AI agent analyzes the conversation and identifies which of the HVE families the
interaction belongs to, which persona archetype the attacker is adopting, which social
engineering techniques are in use, and how deeply the grooming has progressed. This
classification is then matched against the HVE registry to retrieve the named record
with the closest profile, including its pre-scored attack severity metrics that feed the
subsequent scoring stage.

\subsubsection{Victim Profiling (HWE)}

At the same time, the agent extracts signals about the victim's vulnerability state.
These signals are matched against the HWE registry to identify the closest victim
situation. When victim signals are absentas they frequently are, especially in
real-time voice contextsthis stage produces a neutral default.

\subsubsection{Severity Scoring (HVSS)}

The Severity Scoring (HVSS) stage quantifies risk to determine the intervention mode. It
calculates a 0--10 Human Vulnerability Severity Score by combining the Base Score
(attack severity metrics like Impact and Sophistication), derived from the HVE, with the
Context Score (victim psychological status and Match Score), derived from the HWE. The
resulting score and Spell Intensity Flag (Low, Medium, or High) govern Intervention
Selection (HVP), dictating whether the agent employs logic and education or
pacing-and-leading strategies.

\subsubsection{Intervention Selection (HVP)}

The Spell Intensity drives lookup of the appropriate HVP patch for the identified attack
pattern and victim situation. The patch provides structured guidance (verification
steps, post-event steps), an escalation policy, and intervention scripts calibrated to
the computed Spell Intensity (explained in Section~\ref{sec:spell}).

The completed assessment attack family, persona archetype, techniques, HVSS score,
Spell Intensity, and selected intervention guidance is stored per case and surfaced to
the agent's downstream reasoning, which uses it to frame its response and
recommendations.

\subsection{Spell Intensity and Intervention Branching}\label{sec:spell}

Spell Intensity is the operational output of the HVSS computation that governs how the
agent intervenes. The three tiers Low, Medium, and High correspond to qualitatively
distinct psychological states in the victim, and the behavioral literature prescribes
different intervention strategies for each. The mapping from score to strategy is
grounded in dual-process research (Section~\ref{sec:hve}), the hot--cold empathy gap
literature (Loewenstein, 2005), and empirical findings on cognitive friction and
pacing-and-leading.

\subsubsection{Low Spell Intensity: Logic and Education}

At low Spell Intensity, the victim is in a predominantly System-2 cognitive state:
uncertain or curious, but not yet ``emotionally hijacked.'' Analytical reasoning remains
accessible. The appropriate intervention is direct and informational. The agent may
surface explicit pattern-recognition warnings (``This call matches a known authority
impersonation pattern. Police officers do not request wire transfers.'') and ask
disambiguation questions that surface the attacker's tells. Direct contradiction is safe
at Low Spell Intensity because the victim's System 2 is engaged.

\subsubsection{Medium Spell Intensity: Guided Verification}

At Medium Spell Intensity, System 1 is increasingly dominant, but analytical capacity is
not yet fully suppressed. Direct contradiction becomes riskier; it may trigger the
backfire effect. The appropriate intervention shifts to structured verification: the
agent introduces friction via a mandatory pause, frames verification steps as protective
rather than obstructive, and prompts for trusted-contact consultation. A 30-second
mandatory delay, framed as a standard security protocol, provides the temporal window
for partial System 2 recovery.

\subsubsection{High Spell Intensity: Pacing, Leading, and Reality Re-Anchoring}

At High Spell Intensity, the victim is in the full visceral hot state (Loewenstein,
2005) System 1 dominant, System 2 effectively offline. Direct contradiction can
actively deepen the victim's commitment through reactance and confirmation bias. The
appropriate strategy is pacing and leading: the agent matches the victim's emotional
register (pacing), then gradually redirects toward verification behaviors framed as
serving the victim's own stated goal (leading).

\section{HVE Framework Use Cases}\label{sec:usecases}

To effectively demonstrate the value of HVE in each case, this section adopts the
following structured format for each use case: (1) \emph{The Challenge}: a concise
statement of the current operational liability; (2) \emph{HVE Value Proposition}: how
HVE's unique psychological data and scoring resolves this liability; and (3) \emph{Value
Drivers}: the specific HVE features (Codes, HVSS, HVP, Reporting) and their direct,
measurable outcomes (e.g., cross-institution intelligence sharing, loss-reduction
metrics, real-time interruption).

This section outlines the integration of the HVE framework into existing operational
workflows, providing a standardized telemetry layer for high-stakes decision-making
across several critical sectors.

\subsection{Fraud Teams (Analysts and Investigators)}

\textbf{The Challenge:} Traditional fraud engines detect anomalies but struggle with
human-centric frauds such as APP (Authorised Push Payment), where a legitimate user is
manipulated into initiating the transfer.

\textbf{HVE Value Proposition:} HVE shifts fraud defense from reactionary detection to
real-time intervention by enriching transaction signals with human context and a
contextual severity score (HVSS).

\textbf{Value Drivers:}
\begin{itemize}
  \item \textbf{Interoperable HVE Codes:} Standardized codes replace ad-hoc internal
  labels, enabling cross-institution intelligence sharing and proactive ``Vulnerability
  Advisories'' when a code spikes at one institution.
  \item \textbf{HVE as an Enrichment Layer:} A transaction valid via MFA can be enriched
  with telemetry matching an HVE record (e.g., \texttt{HVE-2026-INVT-0042}), triggering a
  high-risk profile (HVSS 8.7 Critical + High Spell).
  \item \textbf{CNA Governance Model:} Borrowing from CVE, authorized entities (banks,
  ISPs) contribute to the global corpus, keeping the taxonomy evolving as fast as
  adversaries.
  \item \textbf{Enterprise Reporting:} Segment heatmaps by channel/region,
  patch-effectiveness measurement (loss-per-user after HVP deployment), and
  time-to-interrupt tracking from grooming to transaction phase.
\end{itemize}

\subsection{Contact Centers and Support Teams}

\textbf{The Challenge:} Support representatives triage cases from free-text narratives
and intuition, producing inconsistent severity assessment and interventions miscalibrated
to the situation.

\textbf{HVE Value Proposition:} HVE converts unstructured customer and member
interactions into structured assessments with intervention strategies calibrated to the
victim's Spell Intensity, replacing gut-feel triage with evidence-based response.

\textbf{Value Drivers:}
\begin{itemize}
  \item \textbf{Spell-Calibrated HVPs:} Intervention prescription is driven by computed
  Spell Intensity logic-and-education at Low, guided verification at Medium,
  pacing-and-leading at High grounded in the hot--cold empathy gap literature.
  \item \textbf{Structured Guidance:} Each HVP carries verification steps, escalation
  policy, and remediation actions across the full Protection--Prevention--Remediation
  life cycle, standardizing handoff between intake, investigation, and recovery.
  \item \textbf{Cross-Center Reporting:} Intervention effectiveness by Spell tier,
  predisposing-factor heatmaps for outreach prioritization, and time-to-stabilization as
  a leading indicator of remediation quality.
\end{itemize}

\subsection{Cybercrime Teams}

\textbf{The Challenge:} Cybercrime professionals and investigators need to identify and
locate the assets and humans behind scam operations, but available signals including
victim narratives, transaction trails, and fragmented reports have limited information
about the attacker.

\textbf{HVE Value Proposition:} HVE allows deploying AI-powered engagements with
attackers and turning those into structured attacker telemetry, capturing a
machine-readable fingerprint of the attacker (Family $\times$ SET $\times$ Persona) that can be matched, clustered, and routed to attribution workflows.
This method was defined by Charm Security as the HoneyBot Operation.

\textbf{Value Drivers:}
\begin{itemize}
  \item \textbf{HoneyBot Operation as HVE Capture:} Decoy interactions generate HVE
  records describing the attacker's signature, converting attacker behavior into the same
  structured telemetry used in victim-facing defense.
  \item \textbf{Vector-String Fingerprinting:} Each engagement produces a compact HVEV1
  vector enabling similarity matching across decoys and jurisdictions, surfacing the same
  operator across campaigns even when cover stories differ.
  \item \textbf{Cross-Framework Interoperability:} HVE-IDs cross-reference F3, ATT\&CK,
  CAPEC, and CVE, joining cognitive-exploit fingerprints with attacker-side
  infrastructure (phishing kits, hosting, payment rails) for attribution inside existing
  SOC tooling.
  \item \textbf{Investigative Reporting:} Operator clustering across deployments,
  channel-family migration as a leading indicator of evasion, and Grooming Depth trends
  signaling attacker sophistication or organized-crime involvement.
\end{itemize}

\subsection{Cybersecurity Awareness Teams}

\textbf{The Challenge:} Security awareness training is ineffective because knowledge
acquired in a calm ``cold'' state does not transfer to behavior in the ``hot'' emotional
state induced by an active scam, and teams lack shared vocabulary, audience-targeting,
and outcome measurement.

\textbf{HVE Value Proposition:} HVE turns education and training from generic ``spot the
phishing email'' exercises into targeted cold-state preparation and/or live
interventions, aligned with the same codes and behaviors that production HVPs deliver in
hot-state moments.

\textbf{Value Drivers:}
\begin{itemize}
  \item \textbf{HVE-Aligned Training:} Modules map to specific HVE families and Social
  Engineering Tactics, giving learners vocabulary that aligns with what their
  organizations use operationally.
  \item \textbf{HWE-Targeted Outreach and Intervention:} Predisposing Factors let teams
  target high-vulnerability segments (e.g., ``Social Isolation with Authority Bias'')
  with the AUTH (Bail/Arrest sub-family) curricula they are most likely to encounter, rather than
  broadcasting undifferentiated content.
  \item \textbf{HVP as Teachable Behavior:} The verification steps, friction practices,
  and pacing-and-leading patterns inside HVPs are themselves the cold-state behaviors
  curricula teach, closing the cold-to-hot transfer gap.
\end{itemize}

\section{Ethics, Privacy, and Responsible Disclosure}\label{sec:ethics}

HVE catalogues the psychological mechanisms attackers exploit and the victim profiles
those mechanisms succeed against. The same description that enables defense also
describes a targeting recipe. The CVE ecosystem confronted an analogous tension and
resolved it through tiered disclosure and federated governance; HVE adopts that posture,
with adjustments for the asymmetric dual-use risk between attack-side and victim-side
records.

\subsection{Dual-Use Risk}

The four registries carry unequal dual-use risk. HVE attack-pattern records describe what
adversaries already know marginal uplift to attackers is small, the defensive value of
a shared vocabulary is large, and the CVE precedent justifies open publication. HWE
records are different in kind: a profile such as ``Social Isolation with Authority Bias''
is simultaneously a defender's outreach signal and an attacker's targeting recipe. HVPs
sit between the two: publishing exact scripts lets attackers script around them, but
obscurity is not a robust defense. The registry's access model reflects this asymmetry
rather than treating the corpus as monolithic.

\subsection{Tiered Access and CNA Governance}

The registry operates four access tiers. The HVE corpus (attack patterns, taxonomy,
vocabularies) is publicly available to authenticated researchers and defenders, on the
CVE model. HWE records, including Predisposing Factors, are gated to vetted operational
defenders banks, telcos, victim support organizations, and law enforcement under
signed CNA-style agreements. Full HVP content is released only to deploying partners.
HVSS aggregate scores are freely citable; component weights and worked examples are CNA
only.

Access is contingent on an acceptable-use agreement that prohibits adverse decisions
against individuals (Section~\ref{sec:fairness}), requires audit-log retention, and
provides for revocation. The registry operator publishes an annual transparency report on
grants, revocations, and confirmed misuse.

\subsection{Privacy by Design}

Three structural commitments govern data handling. First, the framework processes call
audio and message content but strips personally identifying information on
ingest names, account numbers, addresses, and other direct identifiers are removed
before telemetry enters the HVE/HWE pipeline, leaving only the linguistic and behavioral
signal needed for classification. Second, records describe vulnerability \emph{patterns}
rather than individual incidents: one Authority Impersonation bail record, not ten
thousand victim records. Third, Predisposing Factors in HWE are categorical bands (age
band, social isolation indicator, financial literacy band) rather than raw attributes,
and the controlled vocabulary design forecloses free-text victim descriptions.

\subsection{Stigmatization, Profiling, and Fairness}\label{sec:fairness}

Cataloguing victim-side factors is the framework's most contested choice and deserves
explicit defense. In that context, HWE is designed to classify exploitable
susceptibility factors, not people. ``Weakness'' is used in the CWE sense: a structured,
non-judgmental classification of conditions that may be exploited in a specific context,
not a label of personal deficiency or blame. HWE records therefore describe contextual,
situational, behavioral, or cognitive factors that may increase risk in a particular
interaction and help defenders calibrate protective interventions.

To prevent misuse, HWE factors are limited to protective purposes only. They are not
intended for use in credit, insurance, KYC, account eligibility, pricing, or any other
adverse decision about an individual. Protected characteristics such as race, ethnicity,
religion, and sexual orientation are not admitted to the HWE vocabulary. Any admitted
factor must be relevant to intervention design, grounded in behavioral evidence or
observed scam patterns, and subject to governance review for misuse, proxy effects, and
discriminatory outcomes.

\section{Future Work}\label{sec:future}

The HVE framework as presented here is the first version (v1): a structural and
conceptual specification sufficient to catalogue, score, and intervene on the
manipulation patterns observed in current scam ecosystems. Several workstreams are
deliberately scoped out of v1 and tracked for subsequent releases. They divide into three
categories: empirical refinement of the v1 specification, evolution of the framework
itself, and expansion of scope to adjacent attack surfaces.

\subsection{Empirical Validation and Refinement}

The v1 specification establishes the schema, vocabulary, and pipeline architecture; the
empirical work needed to validate and calibrate each component is ongoing and is the
priority for the immediate roadmap.

\textbf{HVSS calibration.} The Base and Context sub-metrics, their weights, and the Spell
Intensity thresholds are presented in v1 with illustrative values. Calibration against a
labeled corpus measuring whether higher HVSS scores predict higher victim loss, longer
time-to-disengagement, and lower intervention success rate is required to fix the
operational thresholds at which partners route to escalated intervention modes. The
validation corpus and methodology will be published alongside HVSS v1.1.

\textbf{Classifier performance and inter-rater agreement.} Family, Persona Archetype, and
Social Engineering Tactics fields are designed to be machine-classifiable from
conversation digests, but the classifier accuracy and human inter-rater agreement that
determine the floor on registry quality have not been published. Future releases will
report IAA on a held-out annotated corpus, with a published target threshold below which a
candidate record is held in review rather than admitted to the registry.

\textbf{HVP effectiveness measurement.} The framework binds each HVE record to an HVP
intervention bundle, but the comparative effectiveness of HVP variants particularly at
High Spell Intensity, where pacing-and-leading scripts are most consequential is an open
empirical question. A measurement protocol covering Time to Interrupt, Time to Disengage,
and intervention-induced backfire rate will accompany future releases and will inform HVP
record updates.

\textbf{Taxonomy residual rate.} The 21-family design target of $<$2\% residual coverage
is presented as a target rather than a measured outcome. Validation against the
partner-contributed case corpus and adjustment of family boundaries or subfamily
vocabularies based on observed residuals is scheduled for future releases.

\subsection{Framework Evolution and Governance}

As the registry accumulates records and CNA partners contribute new patterns, the
operational rules governing taxonomy change require explicit specification in a v2
governance document.
\begin{itemize}
  \item \textbf{Inheritance rules} how subfamilies inherit defaults from their parent
  family (default HVP, allowed Tactics, paired HWE shape).
  \item \textbf{Composition rules} how multifamily records (e.g., ROM + INVT ``pig
  butchering,'' AUTH ``bail with crypto'') are constructed, scored, and queried.
  \item \textbf{Versioning rules} which kinds of changes bump the framework version
  (Tactic vocabulary addition vs.\ family addition vs.\ schema field addition), and how
  downstream consumers interpret version-aware records.
  \item \textbf{Deprecation policy} retiring family codes, Tactic terms, or Persona
  archetypes while preserving historical records.
  \item \textbf{Overlap resolution} for incidents that match multiple families with
  conflicting HVPs: when to load both patches, when to defer to a primary, when to require
  analyst adjudication.
  \item \textbf{Ambiguity handling} for records where annotators legitimately disagree
  on Family, Persona Archetype, or Tactics, including a confidence-score field and
  inter-rater agreement targets.
\end{itemize}

The CNA accreditation process described in Section~\ref{sec:ethics} is operational in
pilot form for v1 partners; v2 will formalize the accreditation criteria, audit cadence,
and revocation procedures, and will publish them as a public governance specification
rather than a partner agreement.

\newpage

\section*{References}
\addcontentsline{toc}{section}{References}
\small

\refitem{Aldawood, H., \& Skinner, G. (2019). Reviewing cyber security social engineering training and awareness programs: pitfalls and ongoing issues. \textit{Future Internet, 11}(3), 73.}
\refitem{Asyali, M., Frank, B., \& H\"olzmer, T. (2026). Fake it till you make it: The psychological and communication tactics behind pig butchering scams. \textit{Journal of Cybersecurity, 12}(1).}
\refitem{Axiom Economics. (2024). \textit{Using behavioural economics to understand and prevent authorised push payment fraud}. UK Payment Systems Regulator.}
\refitem{Beals, M., DeLiema, M., \& Deevy, M. (2015). Framework for a taxonomy of fraud. Financial Fraud Research Center at Stanford Center on Longevity \& FINRA Investor Education Foundation.}
\refitem{Boyle, P. A., Yu, L., Buchman, A. S., \& Bennett, D. A. (2014). Correlates of susceptibility to scams in older adults without dementia. \textit{Journal of Elder Abuse \& Neglect, 26}(2), 107--122.}
\refitem{Brightside AI. (2024). \textit{AI-generated phishing vs.\ human attacks: 2025 risk analysis}.}
\refitem{Cialdini, R. B. (2001). \textit{Influence: Science and Practice} (4th ed.). Allyn \& Bacon.}
\refitem{DeLiema, M. (2023). Financial fraud and deception in aging. \textit{Annual Review of Gerontology and Geriatrics, 43}.}
\refitem{Ferreira, A., \& Teles, S. (2019). Persuasion: How phishing emails can influence users and bypass security measures. \textit{International Journal of Human-Computer Studies, 125}, 19--31.}
\refitem{Gragg, D. (2003). \textit{A multi-level defense against social engineering}. SANS Institute Reading Room.}
\refitem{Hancock, J., \& Tessian. (2020). \textit{The psychology of human error. Stanford University \& Tessian.}}
\refitem{Heiding, F., et al. (2024). Devising and detecting phishing: Large language models vs.\ smaller human models. (AI-generated phishing effectiveness study.)}
\refitem{IBM Global Technology Services. (2014). IBM security services 2014 cyber security intelligence index: Analysis of cyber attack and incident data from IBM's worldwide security operations. IBM Managed Security Services.}
\refitem{James, B. D., Boyle, P. A., \& Bennett, D. A. (2015). Mild cognitive impairment and susceptibility to scams in old age. \textit{Journal of Alzheimer's Disease, 38}(4), 965--976.}
\refitem{Kahneman, D. (2011). \textit{Thinking, Fast and Slow}. Farrar, Straus and Giroux.}
\refitem{Kahneman, D., \& Tversky, A. (1979). Prospect theory: An analysis of decision under risk. \textit{Econometrica, 47}(2), 263--291.}
\refitem{Koning, L., Junger, M., \& Veldkamp, B. (2024). Risk factors for fraud victimization: The role of socio-demographics, personality, mental, general, and cognitive health, activities, and fraud knowledge. \textit{International Review of Victimology, 30}(3), 443--479.}
\refitem{Loewenstein, G. (2005). Hot--cold empathy gaps and medical decision making. \textit{Health Psychology, 24}(4S), S49--S56.}
\refitem{Monta\~nez, R., Golob, E., \& Xu, S. (2020). Human cognition through the lens of social engineering cyberattacks. \textit{Frontiers in Psychology, 11}, 1755.}
\refitem{Mouton, F., et al. (2014). Social engineering attack framework. In \textit{Information Security for South Africa (ISSA)}. IEEE.}
\refitem{Pedersen, K. T., Pepke, L., St\ae rmose, T., Papaioannou, M., Choudhary, G., \& Dragoni, N. (2025). Deepfake-driven social engineering: Threats, detection techniques, and defensive strategies in corporate environments. \textit{Journal of Cybersecurity and Privacy, 5}(2).}
\refitem{Shang, M., Ma, T., Liu, J., Song, W., Liang, Z., Xiao, X., \& Ye, Y. (2025). PsyScam: A benchmark for psychological techniques in real-world scams. In Findings of the Association for Computational Linguistics: EMNLP 2025 (pp. 12623–12637). Association for Computational Linguistics.}
\refitem{Stajano, F., \& Wilson, P. (2011). Understanding scam victims: Seven principles for systems security. \textit{Communications of the ACM, 54}(3), 70--75.}
\refitem{Verizon. (2025). \textit{Data Breach Investigations Report}.}
\refitem{Wang, F., \& Kelsay, J. D. (2025). The prevention of online romance scams using a crime script analysis from the victim's perspective. \textit{International Review of Victimology}.}
\refitem{Which?. (2023). \textit{The psychology of scams: Understanding why consumers fall for APP scams}.}
\refitem{Xu, F., Liu, A., \& Li, X. (2025). Victimization mechanisms and countermeasures in telecom network fraud: A dual-system theoretical perspective. \textit{Frontiers in Psychology, 16}, 1637935.}

\newpage
\appendix
\normalsize
\section{Full Cases Summary \& Comparison}
\begin{figure}[H]
\centering
\includegraphics[width=\linewidth]{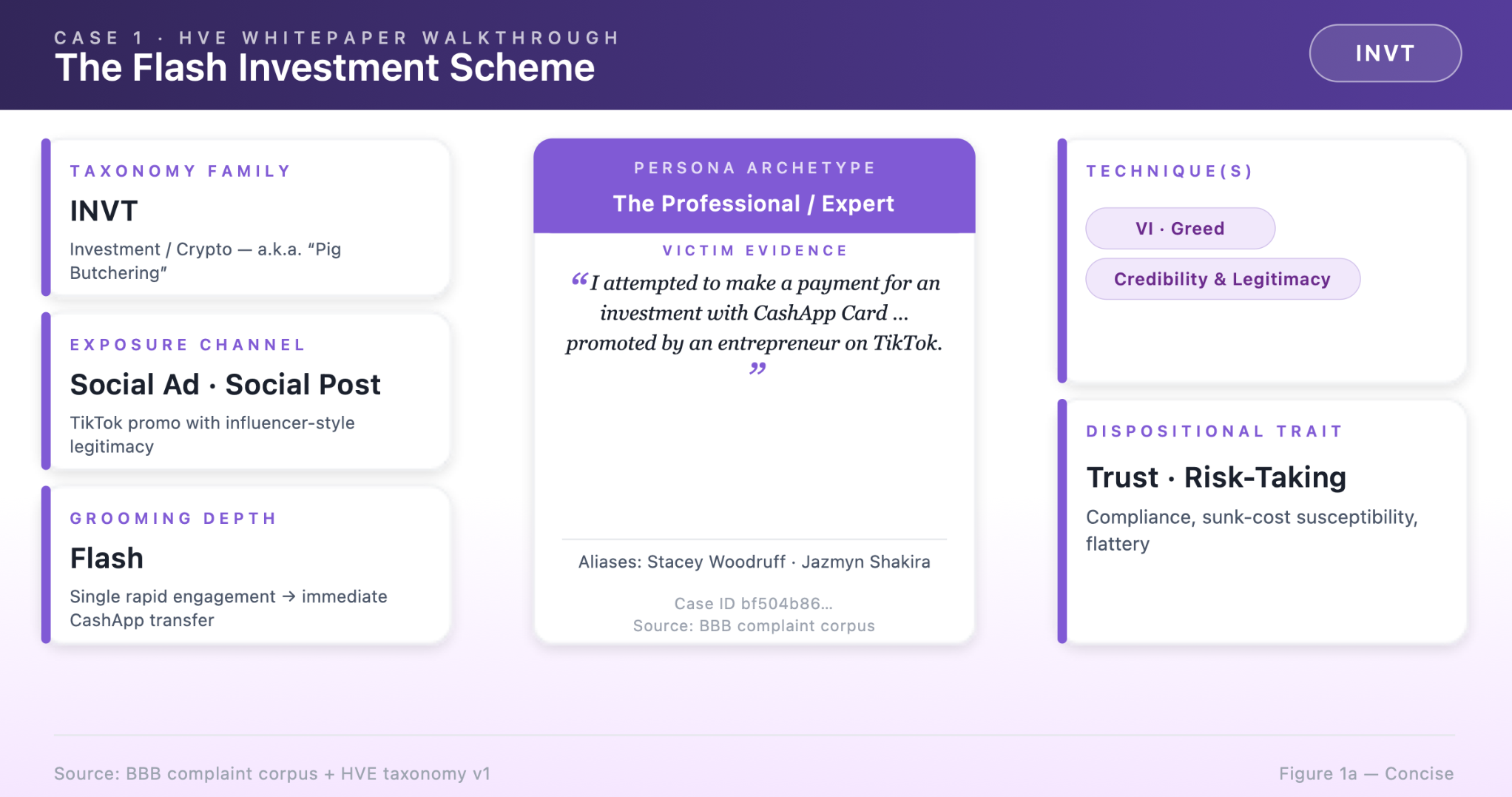}
\caption{HVE Walkthrough of a Flash Investment Fraud}
\label{fig:flash}
\end{figure}

\begin{figure}[H]
\centering
\includegraphics[width=\linewidth]{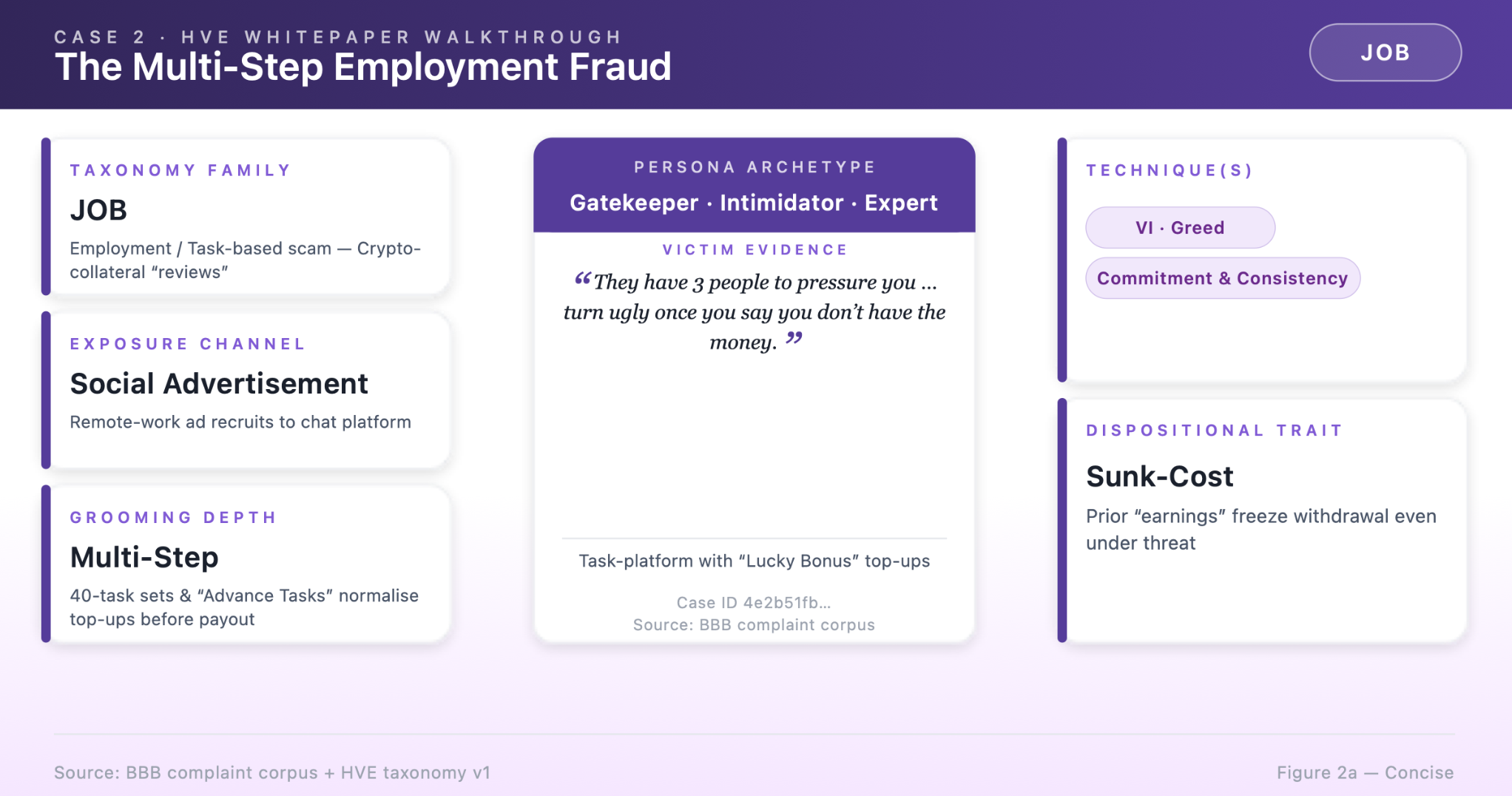}
\caption{HVE walkthrough of a multi-step employment fraud}
\label{fig:multi}
\end{figure}

\begin{figure}[H]
\centering
\includegraphics[width=0.9\linewidth]{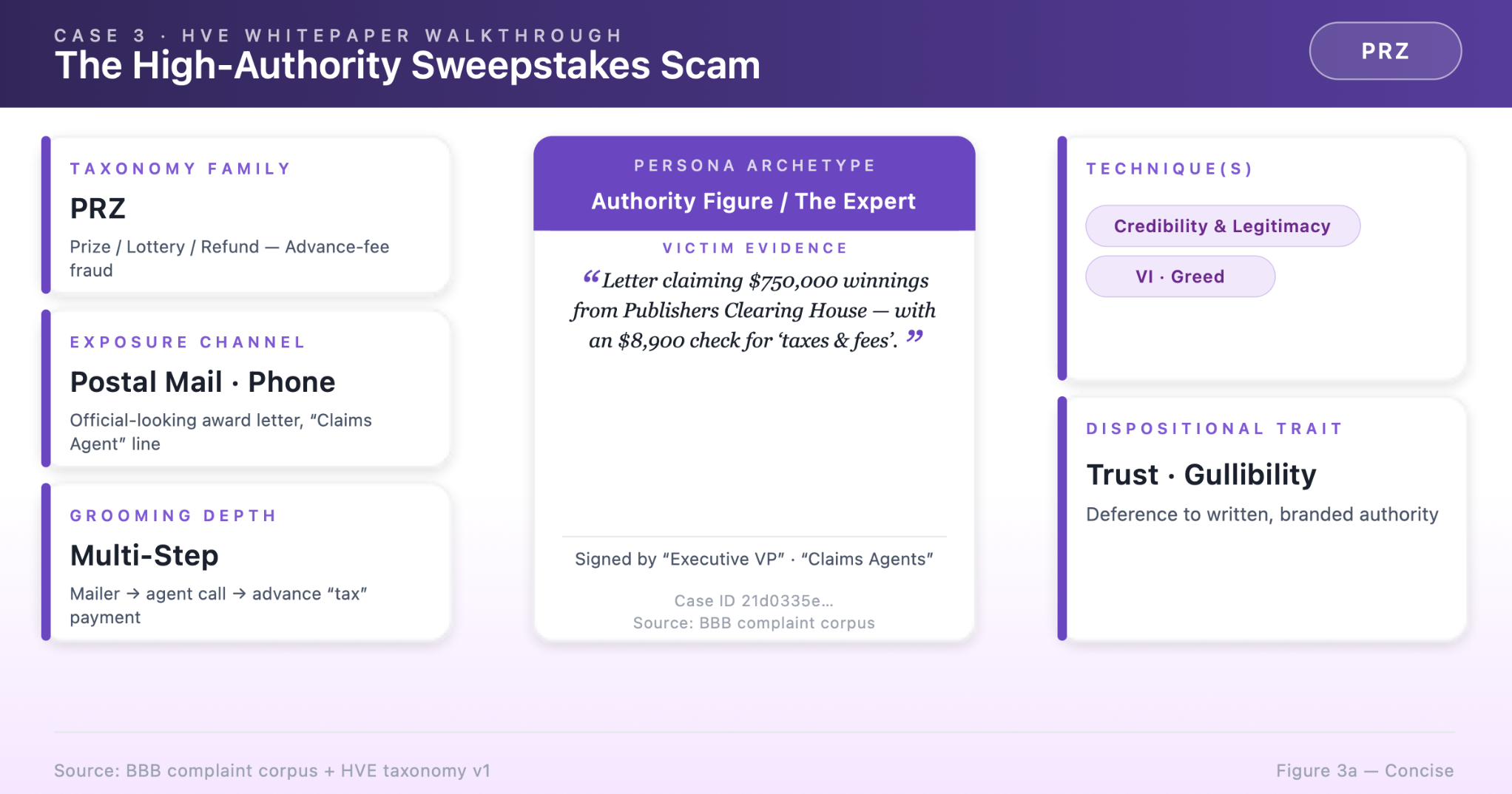}
\caption{HVE walkthrough of a high authority sweepstakes scam}
\label{fig:high}
\end{figure}

\begin{figure}[H]
\centering
\includegraphics[width=0.9\linewidth]{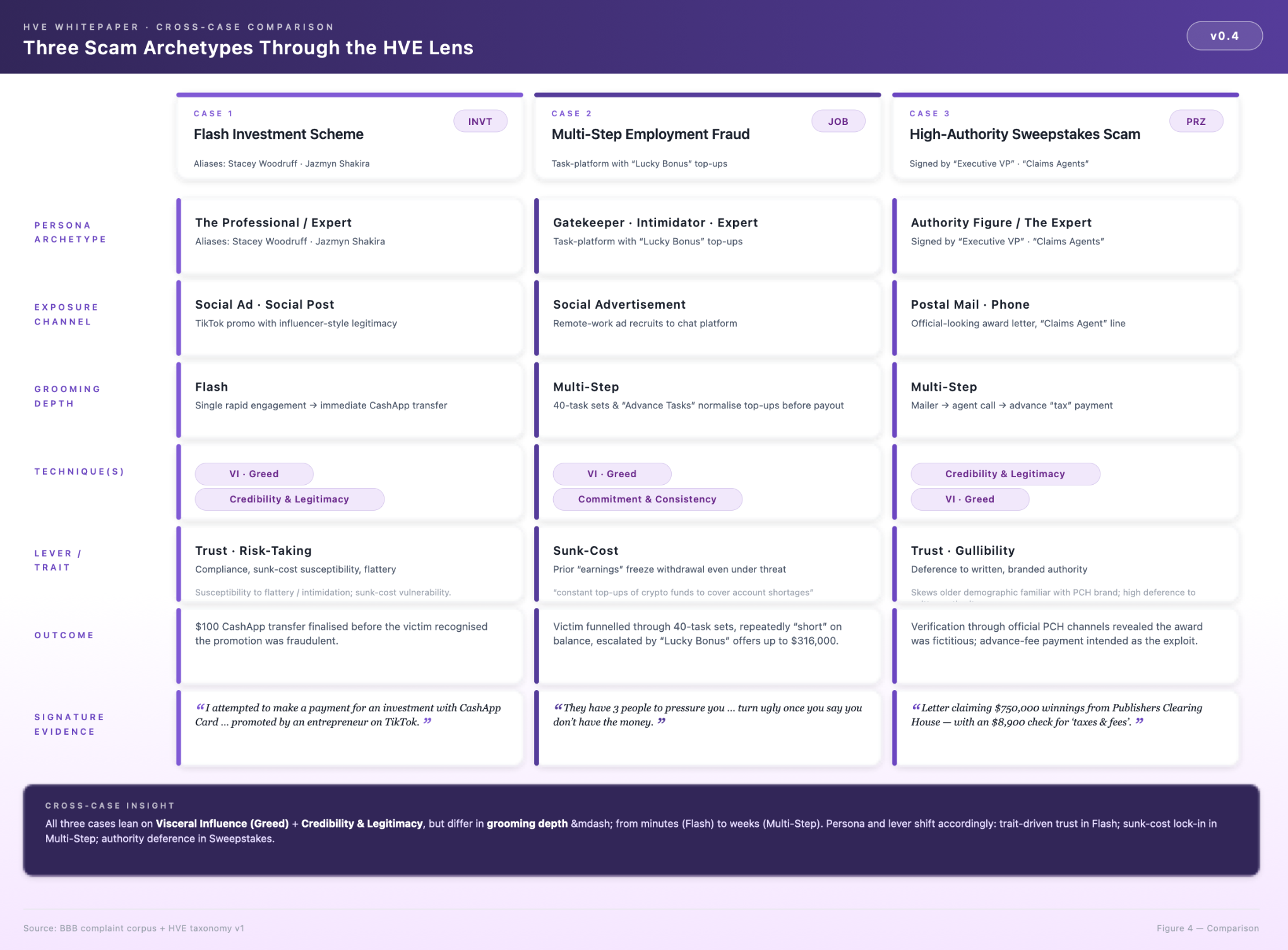}
\caption{HVE high level comparison of the cases}
\label{fig:comparison}
\end{figure}
\newpage
\section{The Full Taxonomy Table}\label{app:taxonomy}

\begin{longtable}{@{}L{0.10\linewidth} L{0.32\linewidth} L{0.50\linewidth}@{}}
\caption{The full HVE Family taxonomy.}\label{tab:taxonomy}\\
\toprule
\textbf{Code} & \textbf{Family} & \textbf{Representative patterns}\\
\midrule
\endfirsthead
\multicolumn{3}{@{}l}{\small\itshape Table~\ref{tab:taxonomy} continued from previous page}\\
\toprule
\textbf{Code} & \textbf{Family} & \textbf{Representative patterns}\\
\midrule
\endhead
\midrule \multicolumn{3}{r@{}}{\small\itshape continued on next page}\\
\endfoot
\bottomrule
\endlastfoot
\textbf{AUTH} & Authority Impersonation & Police, courts, regulators, legal threats, and compliance pressure.\\
\textbf{BANK} & Bank / Financial Institution Impersonation & Fraud-department callbacks, account-locked pretexts, and ``transfer to safe account'' instructions.\\
\textbf{GOV} & Government Services / Tax / Fines & Tax authorities, municipal fines, national-ID issues, and immigration or visa offices.\\
\textbf{TEL} & Telco / SIM / Carrier Impersonation & SIM-swap setup, line-issue pretexts, and verification-code harvesting.\\
\textbf{TECH} & Tech Support / Account Security & Microsoft, Apple, or Google impersonation, remote-access-tool installation, and fake malware alerts.\\
\textbf{ATO} & Account Takeover \& OTP Harvesting & Credential phishing, OTP capture, MFA-push fatigue, and session hijack.\\
\textbf{INVT} & Investment / Crypto / ``Pig Butchering'' & Extended grooming combined with fake trading platforms, escalating deposits, and withdrawal blocks.\\
\textbf{ROM} & Romance / Dating Grooming & Emotional bonding, exclusivity framing, and financial-rescue requests.\\
\textbf{EXT} & Extortion / Blackmail / Sextortion & Threats to leak content, fake legal threats, and coercion loops.\\
\textbf{FAM} & Family Emergency / ``Grandparent'' / Kidnapping Panic & Fake relative-in-trouble calls, deepfake voice or video, and urgent bail, medical, or crisis transfer.\\
\textbf{PRZ} & Prize / Lottery / Giveaway / Refund & Won-prize notifications, refund-available claims, fee-to-claim gating, and card-confirmation pretexts.\\
\textbf{JOB} & Employment / Recruitment / Task Scams & Fake job offers, task platforms with upfront payments, equipment fees, and laundering via payouts.\\
\textbf{MKT} & Marketplace / Escrow / Delivery / Rental & Buyer and seller scams, fake escrow, fake shipping labels, and off-platform payment steering.\\
\textbf{BIZ} & BEC / Invoice / Supplier Payment Diversion & Business email compromise, invoice edits, and banking-details swap on supplier payments.\\
\textbf{EXEC} & Executive / CEO / Internal Impersonation & Confidential-transfer pretexts, CEO gift-card requests, and insider social engineering.\\
\textbf{CHAR} & Charity / Donation / Disaster Relief & Fake NGO pages, empathy and guilt triggers, and quick-donation pressure.\\
\textbf{HLTH} & Healthcare / Insurance / Medical Billing & Fake clinics and insurers, urgent-treatment-payment pretexts, and fraudulent billing corrections.\\
\textbf{UTIL} & Utilities / Subscription Renewal / Shutoff Threats & Shutoff threats for electricity, water, or internet; renewal scams; and immediate-payment pressure.\\
\textbf{IDV} & Identity / KYC / Verification / Document Capture & Verification pretexts, ID or selfie upload demands, and phishing KYC portals.\\
\textbf{LOAN} & Loans / Debt Relief / Credit Repair & Advance-fee lending, debt-settlement scams, and guaranteed-approval framings.\\
\textbf{TRVL} & Travel / Visa / Immigration / Ticketing & Fake tickets and visa services, urgent-document-fee pretexts, and travel-related urgency framings.\\
\end{longtable}

\section{Social Engineering Tactics Factors Table}\label{app:tactics}

\textit{Factor: Social Engineering Tactics - persuasive / psychological techniques
that scammers use to evoke emotions and errors in judgment.}

\begin{longtable}{@{}L{0.26\linewidth} L{0.68\linewidth}@{}}
\caption{Social Engineering Tactics factors.}\label{tab:tactics}\\
\toprule
\textbf{Term} & \textbf{Description}\\
\midrule
\endfirsthead
\multicolumn{2}{@{}l}{\small\itshape Table~\ref{tab:tactics} continued from previous page}\\
\toprule
\textbf{Term} & \textbf{Description}\\
\midrule
\endhead
\midrule \multicolumn{2}{r@{}}{\small\itshape continued on next page}\\
\endfoot
\bottomrule
\endlastfoot
\textbf{Visceral Influence}\newline \textit{Greed, Excitement, Sexual desire, Fear}\newline (and, rarely used by scammers: \textit{Pain, Hunger and thirst}) & Primal drives that bypass rational thinking. Can be short-lived (sudden fear) or longer-term (sexual desire).\\
\textbf{Liking} & Scammers express flattery and liking towards the victim.\\
\textbf{Similarity} & Scammers pretend to be similar to the victim.\\
\textbf{Credibility and legitimacy} & How legitimate, credible, and trustworthy the scam or scammer appears.\\
\textbf{Evoking social norms} & Scammers exploit socially desired norms such as being kind, giving, helpful, etc.\\
\textbf{Authority} & Scammers evoke authority by pretending to be lawyers, government bodies, police, doctors, pastors, etc., or by using known warranties, seals of approval, etc.\\
\textbf{Commitment and consistency} & Scammers evoke commitment through asking for a simple reply, small sums, etc. because they know that once they engage the victim, the victim will continue to respond.\\
\textbf{Scarcity} & When things are scarce, they appear more valuable. Scams often mimic scarcities (e.g., COVID and shortages of masks).\\
\textbf{Urgency} & Rushing the victim or limiting time (e.g., 24 hours to update an account).\\
\textbf{Social proof} & Scammers target affinity groups and use group members to promote the scam (people who socialize and work together), or use fake testimonials to appear as if the offer has the backing of other people.\\
\textbf{Complicity} & Scammers tell the victim they are getting involved in something illegal benefiting the victim (e.g., insider information for trading). This illegal component silences the victim, who may be afraid to report the fraud. (Useful for explaining money mules.)\\
\textbf{Proximity and vividness} & Proximity relates to how proximate the reward is; vividness relates to how enticing it is.\\
\textbf{Priming} \textit{(Semantic, Visual)} & Priming can be semantic (words that bring associations) and visual (logos, colors, etc.); it works at the subliminal level and can influence memory and decisions.\\
\quad \textit{Semantic priming} & Accepting loss (component of financial scams); accepting fate (component of sextortion); to invest (investment scams when the scammer mentions their own investments to plant the idea); to delay (the scammer prepares the victim for a delay in receiving funds or goods, a tactic to delay reporting/detection).\\
\quad \textit{Visual priming} & Familiar logos.\\
\textbf{Altercasting} & The scammer puts the victim in the position of a confidante or friend, so the victim feels obliged to act according to the role; it also gives the illusion of control.\\
\textbf{Grooming}\newline \textit{(Persona construction, Desensitization to financial risk, Critical incident)} & A period of communication in which scammers develop trust, layering manipulative techniques until the victim is deeply invested. \textit{Persona construction:} the scammer builds a persona (e.g., army general, CIA agent) supported by credibility or priming cues (expensive cars, private planes) to depict wealth. \textit{Desensitization to financial risk:} gradually building up to a large amount. \textit{Critical incident:} in romance frauds, after sufficient grooming, a manufactured crisis (medical emergency, collapsing business deal) puts the victim in a state of panic; often the victim offers money rather than being asked, because of the trust built during grooming.\\
\end{longtable}

\end{document}